\documentclass[aps,prd,groupedaddress,nofootinbib,letterpaper]{revtex4}

\usepackage{amsmath}
\usepackage{amsfonts}
\usepackage{amssymb}
\usepackage{amsthm}

\usepackage{braket,mathtools}
\usepackage{mathrsfs}
\usepackage{stmaryrd}

\usepackage{hyperref}
\usepackage{color}

\newcommand{\calo}{{\cal O}}
\newcommand{\calh}{{\cal H}}

\newcommand{\cale}{{\cal E}}
\newcommand{\calp}{{\cal P}}
\newcommand{\calk}{{\cal K}}

\newcommand{\beq}{\begin{equation}}
\newcommand{\eeq}{\end{equation}}
\newcommand{\bea}{\begin{eqnarray}}
\newcommand{\eea}{\end{eqnarray}}

\newcommand{\abar}{{\bar A}}

\begin{document}
\begin{titlepage}

\title{Quantum information/entanglement transfer rates between subsystems}

\author{Steven B. Giddings}
\email{giddings@ucsb.edu}

\author{Massimiliano Rota}
\email{mrota@ucsb.edu}
\affiliation{Department of Physics, University of California, Santa Barbara, CA 93106}

\begin{abstract}
The size of quantum information -- or entanglement -- transfer rates between subsystems is a generic question in problems ranging from decoherence in quantum computation  and sensing, to quantum underpinnings of thermodynamics, to the behavior of quantum black holes.  We investigate such rates for given couplings between subsystems, for sufficiently random subsystem evolution, and find evidence for a conjectured relation of these rates to the size of the couplings.  This provides a direct  connection between entanglement transfer and the microphysical couplings responsible for it.

\end{abstract}

\maketitle

\end{titlepage}

\section{Introduction and motivation}

Consider two quantum subsystems $A$ and $B$ of a quantum system, that are coupled by a Hamiltonian including interactions between $A$ and $B$.  In general, the couplings will transfer quantum information between the subsystems.  This can be characterized in terms of entanglement:  one can take a copy of $A$, maximally entangle it with $A$, and ask how fast the evolution (with trivial evolution of the copy) transfers that entanglement to $B$.  In general, this rate will depend on the interactions, and on the nature of the terms of the Hamiltonian that act solely on $A$ or on $B$.  

This is a very general problem, encountered in various contexts.  One example is that of a quantum computer or sensor interacting with its environment.  In that case, if the interactions transfer entanglement from a subsystem of the quantum system to its environment, this leads to decoherence and degradation of the computing or sensing ability, and indicates how fast the quantum system evolves to classical behavior (see {\it e.g.} \cite{Ahar, Schl} and references therein).  This problem also makes contact with the question of deriving thermodynamics from quantum dynamics\cite{GoEi,GHRRS}.  Specifically, if $A$ and $B$ are large quantum systems, that can for example be thought of as near equilibrium, then the information transfer between them is related to the entropy transfer between them, which may for example be driven by a thermal gradient.  A final example is that of a black hole.  In a quantum field theory description, a black hole will build up entanglement with its environment, either through absorption, or by emitting Hawking particles entangled with its internal state.  But, if a black hole decays and disappears, in order to maintain a quantum-mechanical (unitary) description, this entanglement must be transferred back to its environment during this decay.  An important question is what couplings to the environment could suffice\cite{SGmodels,NVFT,NVU} to transfer sufficient information.  

There are different measures of information transfer and decoherence; this paper will focus on mutual information, which will be used to characterize entanglement transfer.  For large subsystems, coupled by products of operators in the two subsystems, and undergoing sufficiently random evolution, a conjecture about the relation between the information transfer rate and the couplings was made in \cite{NVU}.  This paper will investigate the question of the rate, and provide evidence for this conjecture.\footnote{The related question of bounds on such transfer rates has been studied in \cite{BHV, Brav, AMV}.}  One aspect of our approach is to connect the information transfer to decay probabilities.  Specifically, for subsystem dimensions $|B|\gg|A|$, and if the initial energy of $B$ is comparable or smaller than that of $A$, one typically expects $A$ to decay to its lower-energy states, creating excitations in $B$.  This will be directly related to transfer of entanglement from $A$ to $B$.  Decay probabilities, in turn, can be approximately calculated in perturbation theory, giving a means to estimate information transfer rates.  Alternatively, one can study non-perturbative numerical evolution for certain simple models.  These provide complementary calculations of information transfer rates, agreeing with and extending perturbative results.

\section{Set up and conjecture}

Our goal is to describe how information transfers between two subsystems $A$ and $B$ of a bigger system, whose Hilbert space we assume to be of the form $\calh=\calh_A\otimes\calh_B$.  We take the Hamiltonian to be of the form
\beq\label{Ham}
H=H_A+H_B+H_I\ ,
\eeq
where $H_A$ and $H_B$ act only on their respective subsystems, and $H_I$ is an interaction term which can be expanded in terms of products of operators acting on $A$ and $B$ respectively,
\beq\label{Hint}
H_I= \sum_{\gamma=1}^\chi C_\gamma \calo^\gamma_A \calo^\gamma_B\ ,
\eeq
with coefficients $C_\gamma$.  Transfer of quantum information from $A$ to $B$ can be understood by introducing a copy $\abar$ of $A$, and beginning with a maximally entangled state of $A$ and $\abar$, times some initial state for $B$.  Then, the mutual information $I(A,\abar)$, defined by 
\beq\label{MIdef}
I(A,\abar)=S(A)+S(\abar)-S(A\abar)\ ,
\eeq
with $S(X) = - {\rm Tr}\left( \rho_X \log \rho_X\right)$, 
will initially be $I(A,\abar)=2\log |A|$, and $I(B,\abar)$, similarly defined, will initially be zero.  Evolution via \eqref{Ham} will in general transfer information, alternately understood as  entanglement with $\abar$, from $A$ to $B$, and so in general we expect that initially 
$I(A,\abar)$ decreases and $I(B,\abar)$ grows.  We wish to understand how these changes depend on features of the Hamiltonian \eqref{Ham}.

While this is an interesting general problem, we will impose some additional structure.  We assume that $|B|\gg|A|$, so that $B$ has plenty of ``capacity" to absorb information from $A$, and also that the eigenvalues of $H_A$ have a typical scale $\cale$.  In that case, by energy conservation, evolution by $H$ will typically excite states of $B$ with energy $\cale$ above its initial state -- assuming $H_I$ permits such transitions -- so $\cale$ functions as a common scale and, {\it e.g.}, states of $B$ of much higher energy need not be considered.  In the limit $|B|\gg|A|$, as $A$ decays its number of available states decreases, reducing its capacity for entanglement.
We also define $C_\gamma=c_\gamma \cale$, and so describe $H_I$ in terms of dimensionless coefficients and operators.

A conjecture for the information transfer rate was made in \cite{NVU} in the case where also $|A|\gg1$, where $H_A$ and $H_B$ are sufficiently random, and where $O^\gamma_A$ and $O^\gamma_B$ are  independent (commuting) operators with a standard normalization to unit size.  This conjecture states that for $c_\gamma\lesssim1$, 
\beq\label{conjrate}
\frac{dI(B,\abar)}{dt} = -\frac{dI(A,\abar)}{dt}= \calk \cale\sum_{\gamma=1}^\chi c_\gamma^2\ 
\eeq
with $\calk$  constant. Here we use $I(A,\abar)-S(\abar) = S(A)-S(B) = -[I(B,\abar)-S(\abar)]$; $S(\abar)$ is constant.  For random matrices, \cite{NVU} used operator (largest eigenvalue) norm $\Vert O\Vert=1$, but as is seen below, more generally one should use the norm $|O|^2= {\rm Tr}(O^2)/N$, where $N$ is the subsystem dimension, giving the typical size $\langle O^2\rangle$ of diagonal matrix elements of $O^2$. Information transfers through the $\chi$ different operator ``pathways,"\footnote{A more natural nomenclature might be {\it channels}, but the terminology {\it quantum channel} is already used with a different definition.} corresponding to the independent operators indexed by $\gamma$.  We will also find evidence below that the formula \eqref{conjrate}  holds in a time-averaged sense for small $|A|$.

\section{Perturbative analysis and general treatment}

An intuition behind the preceding conjectures is that, given the assumption $|B|\gg|A|$, and for example the comparability of the sizes of the total energies, system $A$ will decay through $H_I$, in the process transferring entanglement to $B$, and so the rate of transfer of entanglement is related to the decay rate.  We begin by investigating the latter.

Let $|K\rangle$ be energy eigenstates of $H_A$, and $|i\rangle$ of $H_B$; the latter are generically much more closely spaced than the former.  Alternatively, in the case where $H_I$ describes a transition from an initial subset of eigenstates of $A$ to a final subset, and likewise for $B$, these could be states of the subsets; in that case, the final states $|K'\rangle$ of $A$ may belong to a different subset.  Denote the initial state of $B$ as $|\psi_0\rangle$, and for simplicity assume it has energy $E_0$.  Transition rates are given, to first order in $H_I$, by Fermi's Golden Rule.  If $U(t)$ denotes the interaction-picture evolution operator, we find transition amplitudes
\beq\label{pertamp}
\langle i\vert\langle K'\vert U(t)  |K\rangle |\psi_0\rangle = A_{K'i,K0}(t) \simeq -2i \langle i\vert\langle K'| H_I  |K\rangle |\psi_0\rangle \frac{\sin(t\Delta E/2)}{\Delta E} e^{it\Delta E /2}\ ,
\eeq
where 
\beq
\Delta E = E_{K'} + E_i - E_K - E_0\ .
\eeq
The total transition probability from $|K\rangle$ to the collection of final states $|K'\rangle$ is then given by 
\beq\label{Ptrans}
P_K(t) \simeq 4\sum_{K',i} \left|\langle i\vert\langle K'| H_I  |K\rangle|\psi_0\rangle\right|^2 \frac{\sin^2(t\Delta E/2)}{(\Delta E)^2}\ = 2\pi t^2  \langle\psi_0|\langle K| H_I \calp_{1/t} H_I  |K\rangle|\psi_0\rangle
\eeq
where
\beq
\calp_{1/t} =  \sum_{K',i}  |K'\rangle |i\rangle\langle i\vert\langle K'|\ \frac{1}{2\pi} \frac{ \sin^2(t\Delta E/2)}{  (t\Delta E/2)^2}
\eeq
acts like a projector onto states in a band of energies of width $1/t$ about $\Delta E=0$.  
For sufficiently smoothly-varying matrix elements of $H_I$ and density of states, this yields Fermi's Golden Rule, 
\beq\label{FGR}
P_K(t) \simeq 2\pi t|H_I|^2 \rho(E_K+E_0)\ ;
\eeq
the transition rate is approximately constant in $t$, and scales  like the expectation value of $H_I^2$, restricted to the relevant range of states selected by $\calp_{1/t}$.  

If the information transfer rate matched the decay rate, that would demonstrate the essential features of the conjecture \eqref{conjrate}
(the factor of $\cale$ in \eqref{conjrate} arises from the combination of the density of states and the matrix element squared, which both contain the same scale).
However, the former requires closer investigation.  
Specifically, consider the maximally entangled initial state
\beq\label{MII}
|\psi\rangle = \frac{1}{\sqrt N} \sum_{K=1}^N |\bar K\rangle |K\rangle\ ,
\eeq
where $K$ ranges over the $N$ states of either $A$, or of our initial state band of $A$.  The evolution of a given state will be given in terms of amplitudes by
\beq\label{Kevol}
|K\rangle |\psi_0\rangle\rightarrow B_K(t) |K\rangle|\psi_0\rangle + A_{K'i,K0}(t)|K'\rangle |i\rangle\ ,
\eeq
with summation over $i,K'$ implicit, and with
\beq
\left|B_K(t)\right|^2 = 1 - \sum_{K',i} \left|A_{K'i,K0}(t)\right|^2 = 1-P_K(t)\ .
\eeq
The initial state \eqref{MII} then evolves via \eqref{Ham}, \eqref{Kevol} to
\beq
|\Psi(t)\rangle = \frac{1}{\sqrt N} \sum_{K=1}^N |\bar K\rangle \left[B_K(t) |K\rangle|\psi_0\rangle + A_{K'i,K0}(t) |K'\rangle|i\rangle\right]\ ,
\eeq
and the entanglement transfer will be determined in terms of the entropies of the partial traces $\rho_A$ and $\rho_B$.  
These take the form (here we neglect terms where only one of $|\psi_0\rangle$, $|K\rangle$ changes)
\beq\label{rhoA}
\rho_A(t) = \frac{1}{N} \sum_K\left( |B_K(t)|^2 |K\rangle\langle K| + \sum_i |\psi_{iK}(t)\rangle\langle\psi_{iK}(t)|\right)\ ,
\eeq
with 
\beq  
|\psi_{iK}(t)\rangle = \sum_{K'} A_{K'i,K0}(t) |K'\rangle\ ,
\eeq
and
\beq\label{rhoB}
\rho_B(t) = \frac{1}{N} \sum_K\left( |B_K(t)|^2 |\psi_0\rangle\langle \psi_0| + \sum_{K'} |\psi_{K'K}(t)\rangle\langle\psi_{K'K}(t)|\right)\ ,
\eeq
with 
\beq
 |\psi_{K'K}(t)\rangle=\sum_i  A_{K'i,K0}(t) |i\rangle\ .
 \eeq
 The problem is then to relate the resulting difference $S(B)-S(A)$, giving the mutual information, to the decay probabilities $P_K$.

\section{Qubit decay}

As an initial approach to the preceding problem, we consider a simple case, where the system $A$ is a two-state system, {\it i.e.} a qubit.  This will illustrate aspects of the more general problem.

Specifically, suppose $A$ has two states, $|0\rangle$ and $|1\rangle$, which are eigenstates of $H_A$ with eigenvalues zero and $\cale$ respectively.  If $H_I$ connects $|0\rangle$ to $|1\rangle$, such a qubit will generically decay, producing excitation in $B$.  We are interested in evolution of an initial state (see \eqref{MII}, \eqref{Kevol})
\beq
|\Psi(0)\rangle =  \frac{1}{\sqrt2}\left(|\bar0\rangle|0\rangle+|\bar1\rangle|1\rangle\right)\otimes |\psi_0\rangle\ ,
\eeq
and transfer of the entanglement with $\bar A$ from $A$ to $B$.

\subsection{Analytic treatment - simplified model}

For a simplified model of the evolution \eqref{Kevol}, take
\bea\label{simpev}
|1\rangle|\psi_0\rangle & \rightarrow& B(t)  |1\rangle |\psi_0\rangle+ \sum_i A_i(t) |0\rangle|i\rangle\cr
 |0\rangle|\psi_0\rangle&\rightarrow& |0\rangle|\psi_0\rangle \ .
\eea
The decay amplitudes can be perturbatively calculated from \eqref{pertamp}, and satisfy, with $P(t)$ the total decay probability,
\beq
|B(t)|^2 = 1-\sum_i |A_i(t)|^2 = 1-P(t)\ .
\eeq
The appropriate traces of the density matrix arising from the evolution of \eqref{simpev} then give
\beq
\rho_A= \frac{1}{2}\left[1+P(t)\right]|0\rangle\langle0| + \frac{1}{2}\left[1-P(t)\right]|1\rangle\langle1|\ ,
\eeq
\beq
\rho_B=\left[1-\frac{P(t)}{2}\right]|\psi_0\rangle\langle\psi_0| + \frac{1}{2} |\psi(t)\rangle\langle\psi(t)|\ ,
\eeq
with 
\beq
|\psi(t)\rangle= \sum_i A_i(t) |i\rangle\ .
\eeq

These density matrices have von Neumann entropies
\beq\label{Aent}
S(A)=-\frac{1}{2}\left\{\left[1+P(t)\right] \log\left[ \frac{1+P(t)}{2}\right] +\left[1-P(t)\right] \log\left[ \frac{1-P(t)}{2}\right]\right\}
\eeq
and
\beq\label{Bent}
S(B)=-\left[1-\frac{P(t)}{2}\right]\log\left[1-\frac{P(t)}{2}\right] -\frac{1}{2}P(t) \log\left[\frac{P(t)}{2}\right]\ .
\eeq
At $t=0$, $S(A)=\log 2$ and $S(B)=0$; at large time, if $P(t)\rightarrow 1$, $S(A)=0$ and $S(B)=\log2$.  The initial mutual information $I(A,\bar A)=2\log 2$ (see \eqref{MIdef}) thus transfers to $B$, giving final $I(B,\bar A)=2\log 2$.

The perturbative decay probability can be calculated using \eqref{Ptrans}, and for a sufficiently dense collection of allowed final states of $B$, we expect that if $H_I=c\cale O_A O_B$, then $P(t)\propto c^2\cale t$ from \eqref{FGR}.  Perturbation theory fails when $P(t)$ approaches unity, but the perturbative calculation gives the characteristic decay time $T\sim 1/(c^2\cale)$ for it to do so.  Given \eqref{Aent} and \eqref{Bent}, $I(B,\bar A)=S(B)-S(A)+S(\bar A)$ does not vary exactly linearly in time, but does transition in the decay time $T$ from zero to $2\log 2$.  Thus, the conjectured rate \eqref{conjrate} is found in an average over this time.

\subsection{Numerical example}

These statements can be checked beyond a perturbative analysis, through numerical evolution.  Specifically, let the spectrum of $H_A$  be $\{0,1\}$ and consider a simple model where $B$ contains $7$ qubits. For $H_B$ we choose the  spectrum $\{0,1+i/600\}$, where $i$ ranges between $(-63,63)$. After fixing the spectra, the eigenvectors of $H_A$ and $H_B$ are drawn randomly according to the Haar measure. The interaction Hamiltonian is $H_I=c \sigma_A^x\sigma_{B_1}^x$, where $\sigma_{B_1}^x$ acts on one of the qubits in $B$. The non-perturbative numerical evolution can be studied for this model, to check the preceding statements.  For the initial state of $B$ we choose the ground state of $H_B$, $|\psi_0\rangle=|0\rangle$

\begin{figure}[tb]
\centering
{\includegraphics[width=0.5\textwidth]{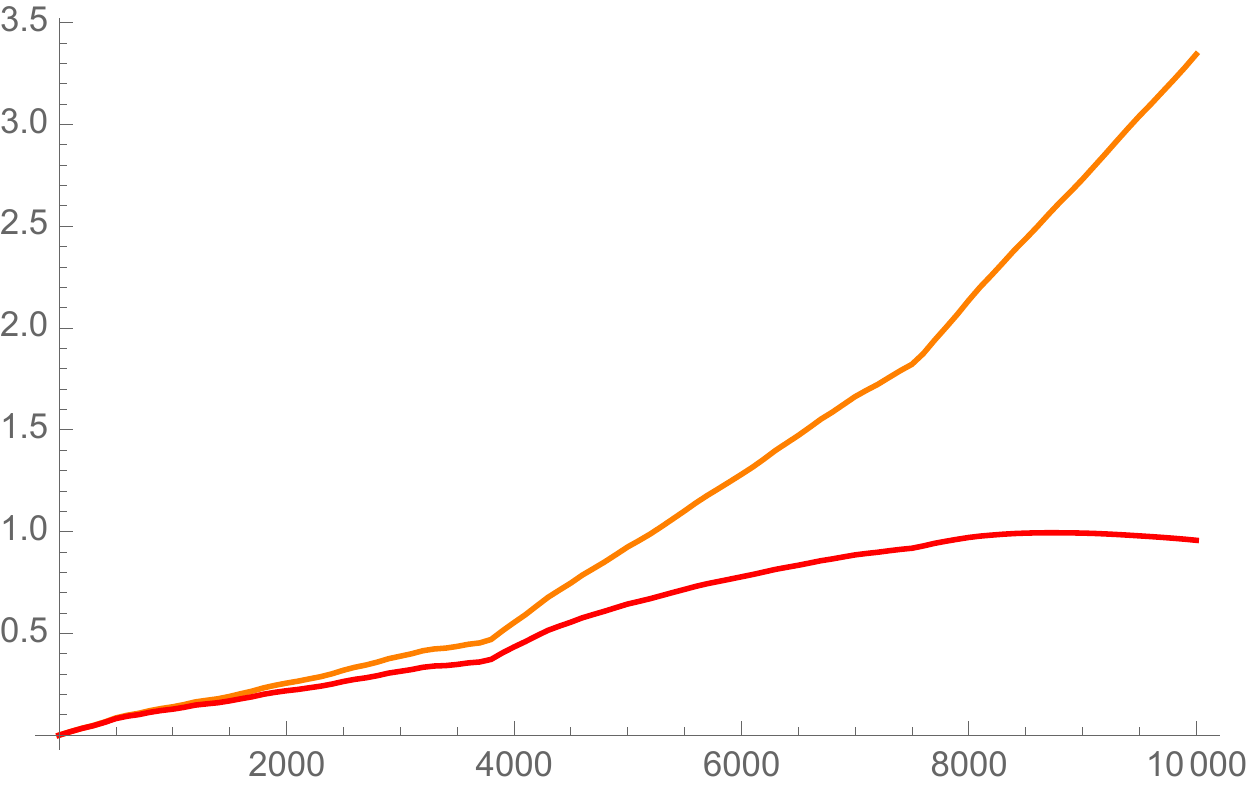}}
\put(8,12){\makebox(0,0){$t$}}
\put(-262,159){\makebox(0,0){$P$}}
\caption{The probability of decay as a function of time, computed numerically (red) and from perturbation theory (orange). The value of the coupling is set to $c=1/400$.}
\label{fig:probability}
\end{figure}

Fig.~\ref{fig:probability} shows the comparison between the perturbative evaluation of the decay probability $P(t)$ and the numerical evolution. One can see that there is good agreement until $P\sim 1/2$,  beyond which the perturbative result gives an increasingly worse approximation. 

\begin{figure}[tb]
\centering
{\includegraphics[width=0.5\textwidth]{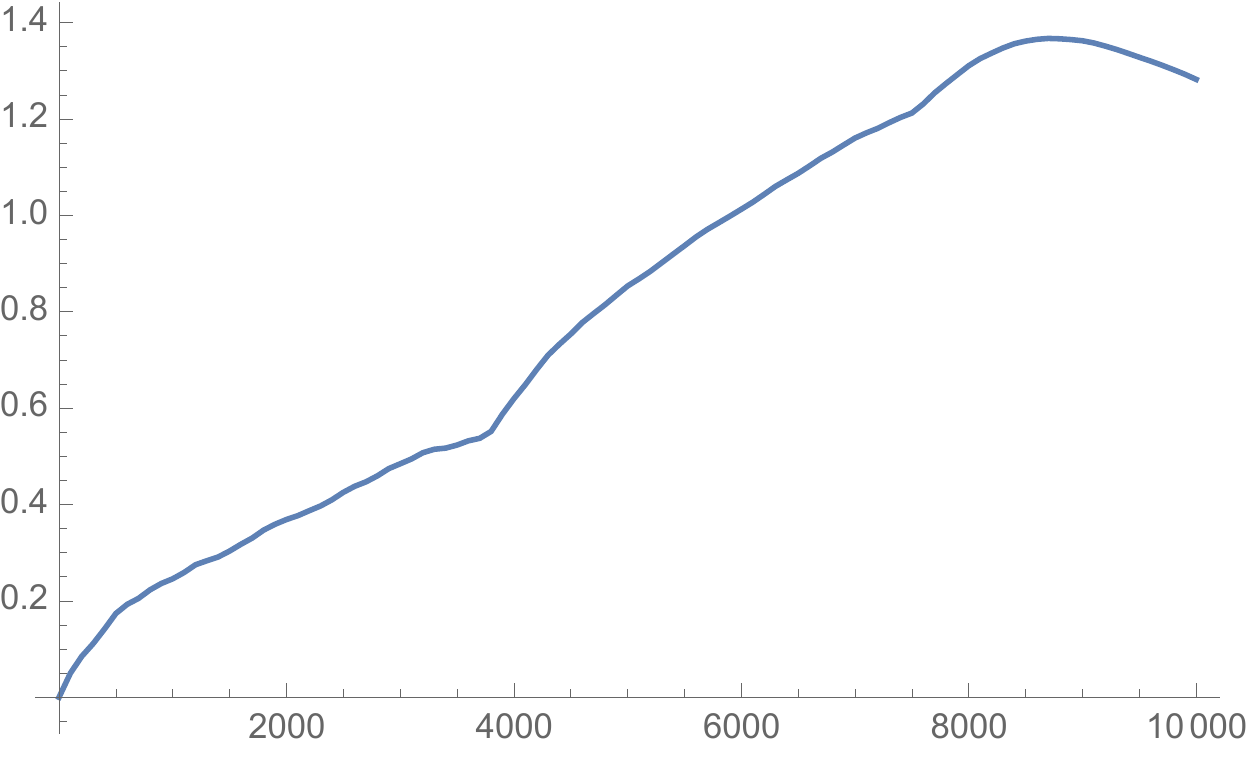}}
\put(8,12){\makebox(0,0){$t$}}
\put(-273,150){\makebox(0,0){$I(B,\bar A)$}}
\caption{The mutual information as a function of time, computed numerically and using the model of the previous section ($c=1/400$). The plot shows perfect agreement, since both curves are plotted but are indistinguishable.}
\label{fig:mutual_information}
\end{figure}

The mutual information $I(B,\bar A; t)$ can be evaluated in two different ways, either directly from the numerically-generated density matrices $\rho_A$ and $\rho_B$, or in the model evolution of the preceding section, with $S(A)$ and $S(B)$ of \eqref{Aent}, \eqref{Bent} evaluated in terms of the numerically-calculated $P(t)$. This model neglects transitions with large energy non-conservation, which should be a good approximation. Indeed fig.~\ref{fig:mutual_information} shows that this gives excellent results for $I(B, \bar A; t)$, since the curves from the two approaches are plotted in that figure and are indistinguishable. Closer examination shows that they differ by very small amounts, $\sim10^{-6}$.

\begin{figure}[tb]
\centering
{\includegraphics[width=0.5\textwidth]{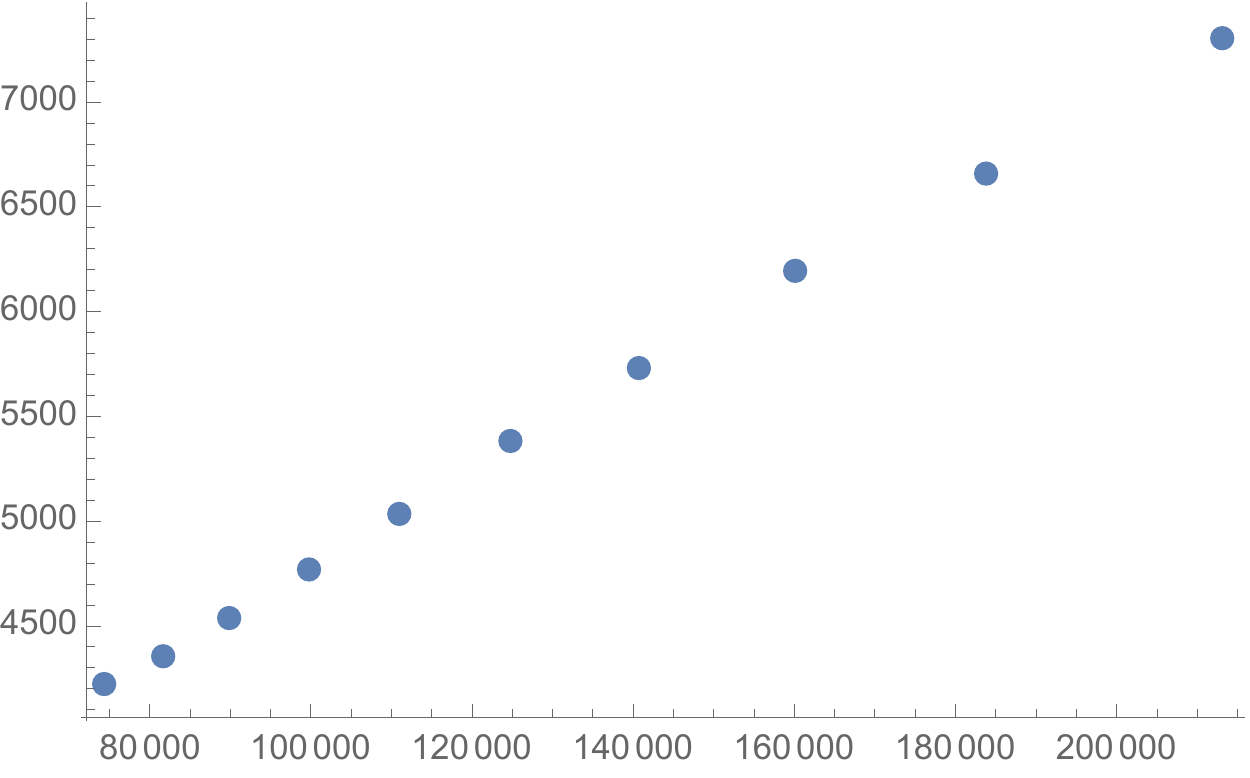}}
\put(15,12){\makebox(0,0){$1/c^2$}}
\put(-260,153){\makebox(0,0){$T_{0.8}$}}
\caption{The dependence of $T_{0.8}$ on the value of the coupling. $c=1/500 + i/6000$ where $i$ ranges between $(1,10)$.}
\label{fig:Tvsc2}
\end{figure}

Finally, the decay time, which Fig~\ref{fig:mutual_information} shows is also the time for $I(B,\bar A)$ to transition from zero to $2\log2$, can also be numerically determined. For example, define $T_{0.8}$ by $P(T_{0.8})=0.8$ as a characteristic decay time.  Fig.~\ref{fig:Tvsc2} shows a plot of $T_{0.8}$ as a function of $1/c^2$, exhibiting the characteristic dependence of the decay time on $c$ that was described above.  So, numerical evolution supports the conjecture, with the proviso that for small $A$ the rate should be viewed as an average information transfer rate.

\section{Next directions}

\subsection{More general analysis}

The preceding provides evidence for the conjecture \eqref{conjrate} in the case where subsystem $A$ is small; we would like to extend that to larger $A$. Specifically, if $A$ has $N$ states in the band of initial states that we consider, and if it transitions to a band of final states with $N'<N$ states, $S(A)$ is expected to decrease by $\log(N/N')$, and $S(B)$ to correspondingly increase.  This will take place over the average transition time, $\bar T$, for the initial state band,  suggesting an information transfer rate
\beq
\frac{dI(B,\abar)}{dt} = -\frac{dI(A,\abar)}{dt} \approx \frac{2\log(N/N')}{\bar T}\ .
\eeq
It is plausible that the averaging over the different initial states, and the different transition times, improves the linear behavior in $t$, and this is in fact seen in simple multistate models\cite{GiRo2}.  An important future goal is to investigate such behavior more closely, for example from the structure of the general formulas $\eqref{rhoA}$, $\eqref{rhoB}$.  Then, from \eqref{Ptrans}, we estimate  
\beq
\frac{1}{\bar T} \approx 2\pi t \langle H_I\calp_{1/t} H_I \rangle \approx k \cale \sum_{\gamma=1}^\chi c_\gamma^2
\eeq
with $k$  a constant. Here we average  over the initial state band; in the second equality, we assume that the operators $O_A^\gamma$,  $O_B^\gamma$ are normalized so that $\langle O^2 \rangle =1$, again for typical allowed (energy-conserving) transitions.  If the $O$ are random matrices, this is the same as saying that the operator norm $\Vert O\Vert\approx 1$, as in \cite{NVU}, though the conjecture for other operators uses the former norm.  Also in the more general case $\log(N/N')\gg1$, we find $\calk=2k\log(N/N')$ in \eqref{conjrate}.

\subsection{Applications}

The preceding discussion is closely related to analogous problems in quantum computing and sensing.  Better understanding the relation of decoherence rates to microphysics may help identify sources of the former.  In fault-tolerant quantum computing (see \cite{NgPr} and references therein), one instead uses a failure measure based on an error rate, though the decoherence rate is in principle related.  Generalization of the present discussion is also needed in that in quantum computing contexts, typically one considers a time-dependent Hamiltonian $H_A(t)$.

One can also anticipate a direct connection to thermodynamics, due to the close connection between thermodynamic entropy and von Neumann entropy.  This paper has investigated transfer of the latter between subsystems, but when these subsystems are behaving thermodynamically, one expects a direct connection to the thermodynamic description of their behavior and entropy transfer.

Finally, the conjecture \eqref{conjrate} was originally motivated by the problem of describing a black hole (BH) interacting with its environment.  This led to an interesting result\cite{NVU}, which can be understood from the conjecture.  In order to transfer information from BH to environment, there must be couplings mediating the transfer, which would be forbidden in a standard local field theory description.  However, to have a close correspondence with field theory, one does not want these couplings to produce large deviations from field theory.  Couplings to degrees of freedom near a BH can be described as interactions of the form \eqref{Hint} between operators acting on the BH and on these nearby field modes.  The couplings can be interpreted in terms of extra fluctuations of the fields near the BH, {\it e.g.} the metric, that depend on the state of the BH.  One finds\cite{NVU} from \eqref{conjrate} that operators of size $|O_{BH}|=1$ transfer sufficient information into modes with BH-size wavelengths.  The BH state space is expected to have very high dimension $N$, {\it e.g.} given by the Bekenstein-Hawking entropy.  The characteristic size of the corresponding fluctuations is given in terms of the expectation value of $O_{BH}$ in a typical state; if the $O_{BH}$'s behave like random matrices, this gives $\langle \psi|O_{BH}|\psi\rangle \sim 1/\sqrt N$ and so this size is very tiny.  One may alternately describe this result in terms of correspondingly tiny couplings between BH energy eigenstates.    In effect, tiny couplings to the huge number of BH states are sufficient to transfer enough information.  This provides a possible way to save quantum mechanics (unitarity) for black holes, with ``minimal" deviation from the quantum field theory approximation, and  impact on infalling observers\cite{NVU}.

\section{Acknowledgements}

We wish to thank 
W. van Dam, A. Jayich, J. Preskill, D. Weld, and especially C. Nayak for helpful discussions.
The work of SG was supported in part by the U.S. DOE under Contract No. 
DE-SC0011702,  Foundational Questions Institute (fqxi.org) 
Grant No. FQXi-RFP-1507, and by the University of California, and that of MR by the Simons Foundation via the ``It from Qubit'' collaboration, and by the University of California.

\bibliographystyle{utphys}
\bibliography{xfer}

\end{document}